\documentclass[12pt]{article}
\usepackage{epsfig}

\parskip=\medskipamount 
plus.3em minus.5em \abovedisplayshortskip=.5em plus.2em minus.4em
\renewcommand{\thesection}{\arabic{section}}

\catcode`@=11

\@addtoreset{equation}{section} \@addtoreset{equation}{subsection}
\def\theequation{\ifnum\value{section}=0 \arabic{equation}\ignorespaces
\else \ifnum\value{section}=-1 A.\arabic{equation}\ignorespaces
\else \ifnum\value{subsection}=0
\thesection.\arabic{equation}\ignorespaces \else
\thesection.\arabic{subsection}.\arabic{equation}\ignorespaces
                             \fi
                        \fi
                   \fi}

{\catcode`\'=\active \def'{{}^\bgroup\prim@s}}

\catcode`@=12

\newcommand{\bq}{\begin{equation}}
\newcommand{\be}{\begin{equation}}
\newcommand{\fq}{\end{equation}}
\newcommand{\ee}{\end{equation}}
\newcommand{\bqr}{\begin{eqnarray}}
\newcommand{\beqs}{\begin{eqnarray}}
\newcommand{\fqr}{\end{eqnarray}}
\newcommand{\eeqs}{\end{eqnarray}}

\def\bop#1{\setbox0=\hbox{$#1M$}\mkern1.5mu
    \vbox{\hrule height0pt depth.04\ht0
    \hbox{\vrule width.04\ht0 height.9\ht0 \kern.9\ht0
    \vrule width.04\ht0}\hrule height.04\ht0}\mkern1.5mu}

\begin{document}
\thispagestyle{empty}

\vskip .6in
\begin{center}

{\bf  String Scale Cosmological Constant}

\vskip .6in

{\bf Gordon Chalmers}
\\[5mm]

{e-mail: gordon$\_$as$\_$number@yahoo.com}

\vskip .5in minus .2in

{\bf Abstract}

\end{center}

The cosmological constant is an unexplained until now phenomena of nature 
that requires an explanation through string effects.  The apparent discrepancy 
between theory and experiment is enourmous and has already been explained 
several times by the author including mechanisms.  In this work the string 
theory theory of abolished string modes is documented and given perturbatively 
to all loop orders.  The holographic underpinning is also exposed.  The 
matching with the data of the LIGO and D0 experiments is also explained to 
the first three or more moments in the cosmological expansion.  
 
\vfill\break

There are three explanations of the cosmological data incorporating the 
energy density, the cold dark matter, galactic curve data and galactic 
halo effect.  The information is encoded in the expansion moment by 
moment in the expansion of the cosmological constant density functional.  
In three prior works a sugra, field theory, and string theory explanation 
have been given \cite{Chalmers1},\cite{Chalmers2}.  These three contexts 
are finally unified in the setting 
of ${\cal N}=2$ sugra and its broken forms in the context of string bits 
which are in fact holographied string fragments.  These string bits have 
never been fully analyzed in the literature except in the context of 
holographic forms and apply in \cite{Chalmers1}.  

The expansion of the exact constant goes like 

\bqr 
V(\Lambda,M) = \sum \Lambda^4 \bigl({\Lambda/M}^2\bigr)^n 
\fqr 
with the first term the discrepancy between the experiment and theory.  
It is commonly assumed that the first two terms dont match and as told 
in \cite{Chalmers1} 

\bqr 
\Lambda^8\over M^4 \ ,
\fqr 
is the measured cosmological energy density to a large factor consisting 
up to nine digits.  This assumes supersymmetry breaking of 3 TeV.  The 
second term namely 

\bqr
{\Lambda^6\over M^2}  \ ,  
\fqr
also does not match with the current experimental data being off by an 
order of 50 digits and indeed is expected to conform to data.  It is explained 
that not only is this term the leading correction to the cosmological 
data from theory.  Both the first $\Lambda^4$ and $\Lambda^6$ will be 
explained in addition to the entire series.  The first two terms are absent.  
The first term absence can be explained by a holographic dimensional analysis 
and the second term by a holographic correspondence with the dimensional 
parameter called the fine structure of matter controlled vacuum.  The third 
moment and its higher order terms generate the cosmological 
expansion with cold dark matter.  The holography is further explained.  

The cosmological constant measures the energy expansion of the universe 
at various stages including the before mentioned effects.  It is difficult 
to resolve in the prior ages due to the absence of a controled expansion.  
One is given here.  Consider string bits which are fragments of a string 
broken into successive halves and quarters never ending.  In actual fact 
there is an experimentally determined minimum length.  They wiggle 
about and actually teleport from one location to the next.  They also 
match holographically transpose with higher dimensional modes.  Take 
the $\Lambda^6/M^2$ term and sum there masses through the formula 

\bqr
\Lambda^4 m^2/M^2 = \sum_{n=0}^\infty \Lambda^4 m^2/M^2 
 e^{- {M/\Lambda}^4 \alpha n} = \Lambda^4 m^2 {\Lambda^4/M^4} \alpha^{-1} 
\fqr 
and agrees with $\Lambda^6/M^2$ when $\alpha=\Lambda^2/M^2$ which is called 
the ambient energy denstiy of spacetime without the string bits.  Furthermore 
there are an infinite number of power order corrections.  

The real constant can now be obtained from the following set $m_0\over\alpha$ 
equal to a prefactor which agrees with the hubble experiment with $\Lambda$ 
the supersymmetry breaking scale of an approximate $2$ TeV.  The string 
scale is $m_0$ and that leads to the ratio of about nine digits.  The 
next term is $\Lambda$ suppressed and generates part of the dark matter 
which is cold for the simple reason that they are string bits.  

Consider ${\cal N}=2$ there is a D-term flatness condition that brings 
about a term which is $\Lambda^6/M^2$ and indeed after summing the entire 
series of the string bits of the gravitational multiplet generate the 
observed value of the cosmological constant.  This is with the fine structure 
constant of matter controled vacuum.  The loop effects have a vertices 
containing four derivatives with the exception of several modes.  The 
string coupling of a massive mode to a massless mode such as the dilatino 
has four derivatives and 

\bqr
{1\over M^2} \int {k_\alpha k_\beta k_\mu k_\nu \over (k^2-m^2)} \ , 
\fqr 
has four derivatives in one gauge and two in the other.  The gauge fixing 
is determined to manifest the coupling dependence and the mass is $m_0 
e^{n \Lambda^4/M^4\alpha}$.  The net result of the integral modewise for 
the entire series of the string bits is controled by $\alpha'$ by the 
expansion previously given with the $\Lambda^4$ term missing.  As explained 
in previous work \cite{Chalmers1} the rest of the higher loop terms are 
suppressed by successive orders $\Lambda^2/M^2$.  The regulator is important 
to determine the supersymmetry broken value of the prefactors.  

The graviton propagator in de Donder gauge has only two derivatives and 
that of the fermionic gravitino has only one and they cancel in softly 
broken supergravity.  In our case we have a graviton which is massive 
by only a tidbit and whose mass is given by a bound state of the massless 
mode in the langrangian together with the hopping of the string tidbit 
from one mode to the other. .  This results in the bound state of the 
graviton between one particle and another with the string tidbit generating 
the mass in the range of 2 TeV.  This is observed in the bifurcation of 
the galactic halo of mercury around the sun in the precession due to the 
bending of light.  Also it occurs as observed in the holographic lensing 
of light around various supernovae including mercury b and supernova a.  
The mass shouldnt be shocking because the matter of gravity hasnt been 
measured in the MeV range.  There are claims to the mass of gravity 
being zero but this isnt clear as there data precludes the precession 
of motion of mercury around the sun in orbital fashion and the two 
degrees of arc motion that general r predicts is wrong.  There are 
another four hundred arcseconds unaccounted for by the holographic 
lensing of matter against light.  This matter is unaccounted for but 
could be made manifest possibly with a four derivative term which 
at low energy really obfuscates the mass.  The mass can be determined 
from that.  It is actually holographic and this is made clear later.  

The entire series 

\bqr 
g = \sum \Lambda^4 \bigl({\Lambda/M}^2\bigr)^n \ , 
\fqr 
sums up to the perturbative contribution.   The holographic contribution 
is written as a double sum with wedges 

\bqr
V(\Lambda,M,\alpha) = f(\Lambda,M,\alpha) \wedge g(\Lambda,M,\alpha) \wedge 
h(\Lambda,M,\alpha) \ , 
\fqr 
which represents multiplication coadjoint like from term to term such as 
the middle representing the perturbative the outer left the holographic 
and the inner right the twist between the two.  It is expected the outer 
right to be a sixth of the inner left due to the conformal group.  In other 
words holography to quantize the form of $f$ to be $f^{1/n \alpha}$. 
The isometry group of $SO(6)$ of the five sphere is expected to play a role 
as in the holographic duality in ${\cal N}=4$ and generates a quantization 
condition in the realistic proposed holography\footnote{In media res work 
not printable yet.}.  The outer 
left is expected to be quantized also.  The two numbers $f^{1/n \alpha}$ 
and $g^{1/m \alpha}$ could be related to the group of $SO(6)$ with 
$n=10$ and the group $D_6$ with twelve generators which twists the 
action with an endomorphism on the spacetime.  The latter is prefered 
for the holographic conjecture and is related to the global structure of 
spacetime.  

The cosmological constant has been explained in view of string bits 
and the holographic contribution clarified.  The holographic extension 
will be exposed in future work on gravity and string theory once finished.  
It is expected to be very accurate to the point of testability in the 
current experiment of LIGO.  Three to six moments are expected to 
be measurable.  

The spectrum of string bits at finite temperature is expected to follow 
from the generating function 

\bqr 
p(n) e^{- k_b T E(n)} \ , 
\fqr 
with $p(n)$ the mode number.  Each particle and string indeed which 
also fit the expansion gives breaks into a tower of bits.  There is 
a possible exception with the tachyon which is not discussed.  The 
probability distribution can be further used at finite temperature 
to mildly put a restriction on the entire series of $f$, $g$, and 
maybe $h$.   

To conclude the data of D0 as described in \cite{Chalmers3} can be used 
to further find the gravitational lensing of the gravitino which is 
claimed to be there and its interaction with matter.  The time spacing 
between the zbgs could show how the cosmological constant appears in 
their experiment.  Two or three measurements of the moments are expected.  

\vfill\break

\end{document}